\documentclass[11pt, letterpaper]{article}

\usepackage[utf8]{inputenc}
\usepackage[T1]{fontenc}
\usepackage{newtxtext,newtxmath} 
\usepackage{microtype} 
\usepackage{amsmath, amsmath, amsfonts}
\usepackage{graphicx}
\usepackage{booktabs}
\usepackage{authblk} 
\usepackage{xcolor}
\usepackage{listings}
\usepackage{geometry}
\usepackage[colorlinks=true, allcolors=blue]{hyperref}

\definecolor{codeblue}{rgb}{0.1, 0.1, 0.8}
\usepackage{listings}
\usepackage{xcolor}
\usepackage{tabularx}
\usepackage{booktabs}
\usepackage{hyperref}

\title{\textbf{Knowledge Graph RAG: Agentic Crawling and Graph Construction in Enterprise Documents}}

\author{Koushik Chakraborty}
\author{Koyel Guha}
\affil{Google AI, Global Services Delivery}
\date{March 30, 2026}

\usepackage{graphicx}
\usepackage{caption}
\usepackage{float}
\begin{document}

\maketitle

\begin{abstract}
This research paper addresses the limitations of semantic search in complex enterprise document ecosystems. Traditional RAG pipelines often fail to capture hierarchical and interconnected information, leading to retrieval inaccuracies. We propose Agentic Knowledge Graphs featuring Recursive Crawling as a robust solution for navigating superseding logic and multi-hop references. Our benchmark evaluation using the Code of Federal Regulations (CFR) demonstrates that this Knowledge Graph-enhanced approach achieves a 70\% accuracy improvement over standard vector-based RAG systems, providing exhaustive and precise answers for complex regulatory queries.
\end{abstract}

\section{Introduction}
In massive enterprise document ecosystems—legal, construction, specifications—information is not flat. It is deeply hierarchical and interconnected. A cohesive "answer" to a query is rarely found in a single text chunk. Instead, it is scattered across a "Base Contract," multiple "Addendum," "Technical Manuals," and "Amendments." Standard RAG pipelines, which depend on semantic similarity, are fundamentally inadequate in this domain. These systems retrieve text based on the most similar content rather than the referenced content. Furthermore, they fail to account for the temporal precedence of recent amendments over earlier versions. To solve this, we must build Multi Agent Workflows with Complex Graph Data Store that act as graph traverser: crawling precise citations and constructing a knowledge graph of validity.

The core contributions of this paper are:
\begin{enumerate}
    \item \textbf{The Agentic Knowledge Graph (AKG) Framework}: A hybrid RAG architecture that moves beyond semantic retrieval to incorporate deterministic graph traversal.
    \item \textbf{The Temporal Graph Schema}: Modeling relationships using explicit SUPERSEDES and REFERS\_TO edges to resolve versioning and multi-hop dependency conflicts.
    \item \textbf{The Recursive Reference Crawler}: An autonomous agent implementation that programmatically follows citation paths to assemble a complete, contextually valid answer.
    \item \textbf{Quantitative Validation}: Demonstration of a 70\% accuracy improvement over vector-only RAG systems in the complex Code of Federal Regulations (CFR) domain.
\end{enumerate}

This deep dive focuses on the two core engineering pillars of this architecture:
\begin{enumerate}
    \item \textbf{The Recursive Reference Crawler}: Identifying and fetching linked nodes.
    \item \textbf{The Knowledge Graph}: Structuring these nodes for intelligent retrieval.
\end{enumerate}

\subsection{Sample documents:}
\subsubsection{Document 1: The Base Contract} 
\textbf{Document ID:} Base\_Contract\_Vol1 \\
\textbf{Date:} 2020-01-01 \\
\textbf{Description: }The foundational construction agreement. \\
\textbf{[Content] Section 4: Structural Materials \& Methods Clause 4.2} All permanent concrete structures, including retaining walls and station boxes, shall utilize Grade 25 Concrete, unless otherwise specified. For curing times, refer to Clause 4.5. \\

\textbf{Clause 4.5} Concrete curing times shall be a minimum of 14 days under standard atmospheric conditions.\\

\textbf{Clause 4.8 }Earth Retaining Stabilizing Structures (ERSS) shall be constructed using standard contiguous bored pile (CBP) methods. No specific integration is required between the main station box and secondary entrances unless directed by the Chief Engineer.\\

\textbf{Section 9: Particular Specifications Clause 9.3.1} General boundary walls shall have a minimum thickness of 600mm.\\

\subsubsection{Document 2: Amendment 01} 
\textbf{Document ID:} Amendment\_01\_Vol2 \\
\textbf{Date: }2022-06-15 \\
\textbf{Description:} First major revision due to updated geological surveys.\\
\textbf{[Content] Item 1: Revision to Materials} Delete Clause 4.2 of the Base Contract (Base\_Contract\_Vol1) in its entirety and replace it with the following: Revised Clause 4.2: All permanent underground concrete structures shall utilize Grade 30 Waterproof Concrete. \\

\textbf{Item 2: Revision to ERSS Methods} Amend Clause 4.8 of the Base Contract. Add the following subsection: Clause 4.8 (a): For high-water-table areas, the ERSS method must deviate from standard CBP. Refer to the updated requirements in Particular Specification Clause 9.3.5. \\

\textbf{Item 3: New Specifications }Add new Clause 9.3.5: Where water tables are within 2 meters of the surface, secant pile walls shall be used in accordance with environmental safety guidelines.\\

\subsubsection{Document 3: Tender Addendum No. 03} 
\textbf{Document ID: }Tender\_Addendum\_03\_Vol1to6 \\
\textbf{Date:} 2024-10-10 \\
\textbf{Description:} Final critical updates right before tender submission. Incorporates user-clarification requests. \\
\textbf{[Content] Clarification No. 12: Concrete Grades} Further to the revisions in Amendment\_01\_Vol2, delete the revised Clause 4.2 and replace with the following: Final Clause 4.2: Due to new structural loads, all permanent station box structures must utilize Grade 40 Concrete (High Strength). Retaining walls not attached to the station box may remain at Grade 30. \\

\textbf{Clarification No. 45: Station Box ERSS \& Entrances} Delete Clause 4.8(a) as introduced in \textbf{Amendment\_01\_Vol2}. Add new Clause 4.8(g) to the Main Text: Clause 4.8(g): Diaphragm walls shall be adopted for all permanent Earth Retaining Stabilizing Structures (ERSS) of the station box. For the associated structural requirements, refer to Clause 9.3.10.5 of the Particular Specification. \\

\textbf{Clarification No. 46: Integration of Entrances} Add new Clause 9.3.10.5 to the Particular Specification: Clause 9.3.10.5: Diaphragm wall thickness for the station box shall be a minimum of 1200mm. Entrances 1 \& 3 are integrated with the structure of the station box and shall be treated as part of the station box for structural purposes. For a clear demarcation sketch of boundaries between station box structures and Entrance structures in plan \& section, refer to \textbf{Drawing\_17.3.1\_Demarcation\_Plan}.

\subsection{Why RAG on these documents are tricky}
Standard RAG pipelines rely on Cosine Similarity to measure semantic relevance between a query vector $\mathbf{A}$ and a document chunk vector $\mathbf{B}$:

\begin{equation}
\cos(\theta) = \frac{\mathbf{A} \cdot \mathbf{B}}{\|\mathbf{A}\| \|\mathbf{B}\|}
\end{equation}

While effective for thematic matches, this metric is fundamentally incapable of discerning the temporal validity or structural context of the information. The graph-based approach presented here shifts the retrieval foundation from this probabilistic measure to a deterministic, set-theoretic traversal logic.

\subsubsection{Example 1: The Cascading Amendment Problem} 
\textbf{User Query:} "What grade of concrete must I use for the permanent station box?"\\

\textbf{The Correct Answer: }Grade 40 Concrete (High Strength). Retaining walls not attached to the station box may remain at Grade 30.\\

\textbf{Why this is correct (and the complexity behind it):} To arrive at this correct answer, a human reader (or a system) must trace a multi-year chain of deletions and replacements across three separate documents:\\

\begin{enumerate}
    \item They must first find \textbf{Clause 4.2} in the Base Contract (2020), which states \textbf{Grade 25} should be used.
    \item They must then realize that two years later, Amendment 01 (2022) legally deleted the original Clause 4.2 and replaced it with a requirement for \textbf{Grade 30}.
    \item Finally, they must discover that right before submission, Tender Addendum 03 (2024) explicitly deleted the revision from Amendment 01, establishing \textbf{Grade 40} as the final, binding requirement for the station box.
\end{enumerate}

\textit{\textbf{The Standard RAG Failure:}} A standard semantic search engine will simply return all three text chunks because they all highly match the keywords "grade of concrete permanent station box". It has no way of knowing that the 2024 document invalidates the 2022 and 2020 documents, practically guaranteeing a hallucinated or conflicting answer.

\subsubsection{Example 2: The Multi-Hop Breadcrumb Trail} 
\textbf{User Query:} "How should we construct the ERSS for the station box, and where can I find the exact boundary measurements for Entrances 1 and 3?" 

\textbf{The Correct Answer:} The ERSS for the station box must be constructed using \textbf{Diaphragm walls} with a minimum thickness of 1200mm. Because Entrances 1 \& 3 are integrated with the station box, the exact boundary measurements and demarcation lines can be found in \textbf{Drawing\_17.3.1\_Demarcation\_Plan}.

\textbf{Why this is correct (and the complexity behind it):} Answering this question requires navigating a convoluted, multi-step breadcrumb trail where no single document or clause contains the complete answer:

\begin{enumerate}
    \item First, the reader must bypass outdated ERSS rules in the Base Contract and Amendment 01, eventually finding the currently valid \textbf{Clause 4.8(g)} in Tender Addendum 03.
    \item Clause 4.8(g) reveals the method (Diaphragm walls) but does not answer the second part of the user's question about entrances. Instead, it provides an instruction: "\textit{refer to Clause 9.3.10.5}".
    \item The reader must then hunt down \textbf{Clause 9.3.10.5}. This clause explains that the entrances are structurally integrated with the station box, but still doesn't give the exact measurements.
    \item Instead, Clause 9.3.10.5 provides a final instruction: look at \textbf{Drawing 17.3.1}.
\end{enumerate}

\textbf{\textit{The Standard RAG Failure:}} If a system only retrieves the most "semantically similar" paragraphs to the user's prompt, it will likely just pull Clause 4.8(g). It will stop there, completely missing the structural requirements in 9.3.10.5 and failing to tell the user which drawing holds the actual measurements. The system must know how to "read" an instruction in the text and physically jump to the next location to assemble the full picture.

\section{Related Work} 
The field of Retrieval-Augmented Generation (RAG) is predominantly anchored in vector space models, leveraging semantic similarity for chunk retrieval. This paper directly addresses the limitations of standard RAG by introducing structural awareness. 

\subsection{Limitations of Standard Vector RAG} 
Traditional RAG systems, which rely solely on vector-based retrieval (such as those using \textbf{Cosine Similarity}), treat documents as an amorphous collection of text chunks. This architecture fails when:
\begin{enumerate}
    \item \textbf{Temporal Conflicts:} A semantically dense, but outdated, clause is retrieved over a concise, superseding amendment (the "temporal hallucination" problem).
    \item \textbf{Contextual Fragmentation:} A complete answer requires explicitly following a chain of citations (multi-hop traversal), which vector search cannot perform deterministically.
\end{enumerate}

\subsection{Advancement over GraphRAG Frameworks}
Our proposed framework—which focuses on \textbf{Temporal-Hierarchical RAG}—advances existing Graph RAG concepts by introducing specific temporal and logical edge types. While many GraphRAG solutions model general relationships, our system prioritizes the engineering of \textbf{SUPERSEDES} and \textbf{REFERS\_TO} edges to enforce document validity and structural dependency, which are crucial for complex, regulated domains. The core innovation lies in using the Knowledge Graph not just for enriched vector indexing, but as a primary, deterministic traversal mechanism guided by an autonomous agent (the Recursive Reference Crawler).

\section{System Architecture and Methodology}
The Reference Crawler is an autonomous agent designed to navigate the "Superseding Logic" of enterprise documentation. Unlike a probabilistic search engine, it follows explicit instructions found in the text (e.g., "\textit{Delete Clause 12.1 and see Amendment 3}").

It operates on a cycle of \textbf{Extraction -> Traversal -> Aggregation.}

\subsection{Core Architecture}
The crawler uses a queue-based traversal algorithm (BFS/DFS) to follow citations (Edges) from one Clause (Node) to another. It uses a specialized LLM call to extract structured citations from unstructured legal text.

\begin{figure}[H]
    \centering
    \includegraphics[width=0.9\textwidth, trim=0 20 0 0, clip]{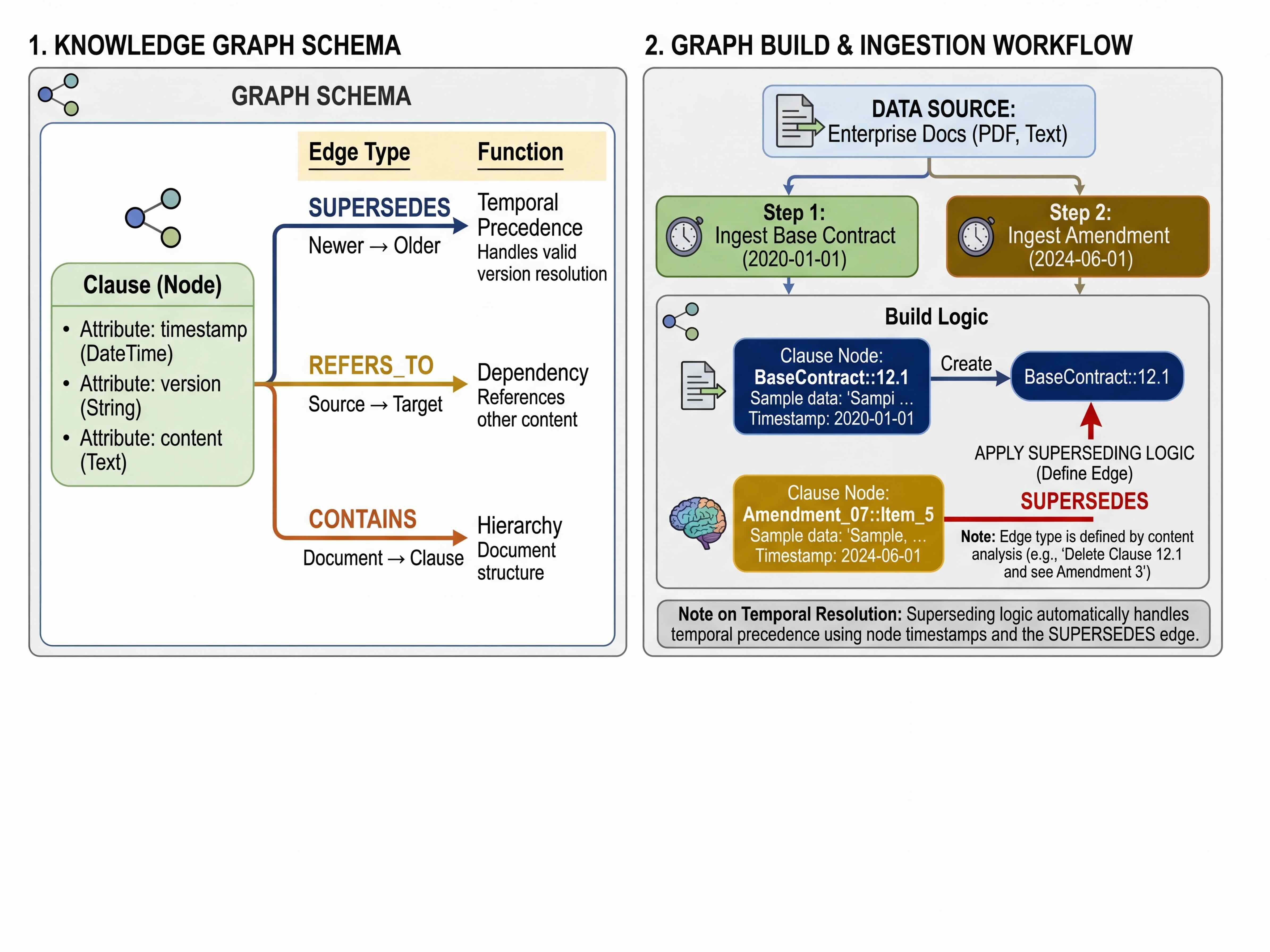} 
    \vspace{-80pt}  
    \caption{Building Knowledge Graph based on the documents}
    \vspace{-10pt}
    \label{fig:knowledge-graph} 
\end{figure}

\begin{figure}[H]
    \centering
    \includegraphics[width=0.9\textwidth, trim=0 20 0 0, clip]{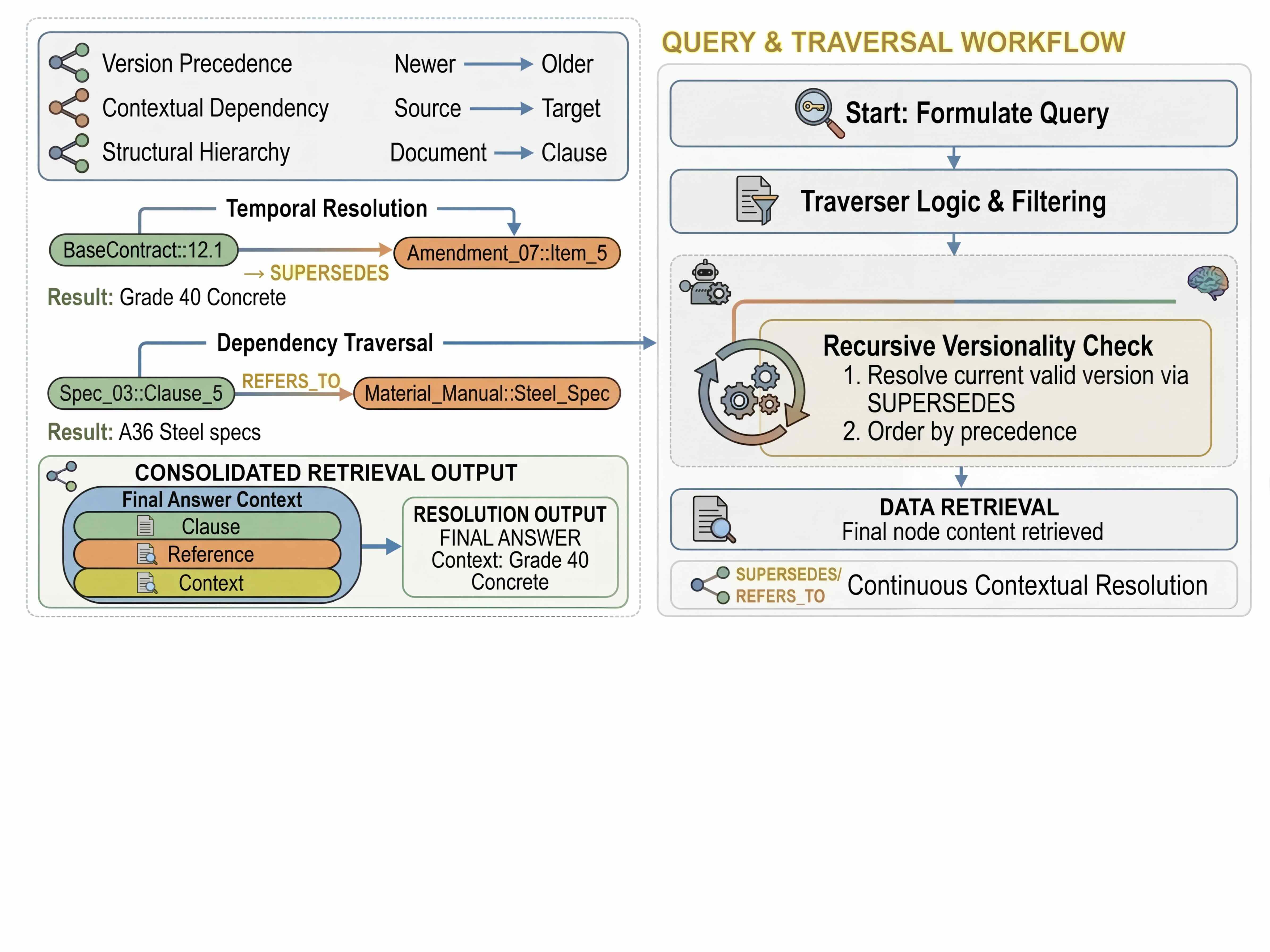} 
    \vspace{-80pt}  
    \caption{Query Processing \& Retrieval}
    \label{fig:processing-retrieval} 
\end{figure}

\subsection{Recursive Crawler Implementation}
Please refer to Appendix for further details.

\section{Building the Knowledge Graph}

While the crawler navigates primarily at query-time, we can optimize global retrieval by pre-building a \textbf{Knowledge Graph}. This graph explicitly models the complex relationships between documents, allowing us to query for "The Valid Clause" rather than just "The Text."

\subsection{Graph Schema}
\begin{itemize}
    \item \textbf{Nodes:} Documents/Clauses (Attributes: timestamp, version, text\_content) 
\end{itemize}
\begin{itemize}
    \item \textbf{Edges:}
\end{itemize}
          \begin{enumerate}
              \item \textbf{SUPERSEDES:} Directional edge from a newer clause to an older one (superseding).
              \item \textbf{REFERS\_TO:} Citational edge indicating dependency.
              \item \textbf{CONTAINS:} Hierarchical edge (Document contains Clause).
          \end{enumerate}

\subsection{Graph Construction}
Please refer to Appendix for further details.

Here are examples of how user questions will interact with the graph built from the three documents given in section \textbf{1.1 Sample Documents}:

\subsubsection{Query Example 1: Testing Temporal Precedence (The "Supersedes" Edge)}
\textbf{User Query: } \textit{"What grade of concrete must I use for the permanent station box?"} \\
\textbf{Correct Answer:} \textit{should return} Grade 40 as the legally valid answer, ignoring Grades 25 and 30. \\

\textbf{Expected Graph Traversal:}

\begin{enumerate}
    \item Standard Search hits \textit{Base\_Contract\_Vol1::Clause 4.2} (Says Grade 25).
    \item Graph checks for \textit{SUPERSEDES} edges targeting this node.
    \item Redirects to \textit{Amendment\_01\_Vol2::Item 1} (Says Grade 30).
    \item Graph checks for \textit{SUPERSEDES} edges targeting this node.
    \item Redirects to \textit{Tender\_Addendum\_03\_Vol1to6::Clarification No. 12} (Says Grade 40).
    \item \textbf{Crawler Output:} Returns Grade 40 as the legally valid answer, ignoring Grades 25 and 30.
\end{enumerate}

\subsubsection{Query Example 2: Testing Recursive Traversal (The "Refers\_To" Edge)}
\textbf{User Query: } \textit{"How should we construct the ERSS for the station box, and where can I find the exact boundary measurements for Entrances 1 and 3?"} \\

\textbf{Expected Graph Traversal:}

\begin{enumerate}
    \item Standard Search hits \textit{Tender\_Addendum\_03\_Vol1to6::Clause 4.8(g)} (Identifies Diaphragm walls are needed).
    \item The Crawler's LLM extracts the reference: \textit{target\_document: Tender\_Addendum\_03, target\_section: 9.3.10.5}.
    \item Crawler hops to \textit{Clause 9.3.10.5} (Identifies 1200mm thickness and that Entrances 1 \& 3 are integrated).
    \item The Crawler's LLM extracts the next reference: \textit{target\_document: Drawing\_17.3.1\_Demarcation\_Plan.}
    \item \textbf{Crawler Output:} The LLM aggregates all three hops into a single context window. It answers the ERSS method (Diaphragm walls), the thickness (1200mm), notes the integration of Entrances 1 \& 3, and successfully directs the user to \textit{Drawing 17.3.1} for the exact boundary plan—a complete, hallucination-free answer that a standard semantic search would miss.
\end{enumerate}

\section{Benchmark Performance Evaluation}
\subsection{Dataset Profile: The Code of Federal Regulations (CFR)}
The benchmark utilizes the \textbf{Code of Federal Regulations (CFR)}, sourced in structured XML format from \href{https://www.govinfo.gov/bulkdata/CFR}{https://www.govinfo.gov/bulkdata/CFR}. The CFR was selected as the primary corpus because its structural complexity serves as a rigorous stress test for information retrieval systems. Specifically, it exhibits two characteristics that highlight the limitations of flat vector stores compared to Knowledge Graphs (KGs):
\begin{itemize}
    \item \textbf{Inherent Hierarchy:} The CFR follows a strictly nested taxonomy, descending from \textit{Title through Chapter, Subchapter, Part, Subpart, and Section, down to individual Paragraphs.}
    \item \textbf{Dense Cross-Referencing:} Regulatory language is characterized by frequent explicit citations (e.g., "pursuant to 261.14(a)(4)"), creating a complex web of interdependent legal authorities.
\end{itemize}

\subsection{Methodology and Accuracy Results}
To evaluate performance, we curated a gold-standard set of 20 complex regulatory questions. We compared the retrieval accuracy of a \textbf{Knowledge Graph-enhanced approach} against a standard \textbf{Vector-based Retrieval-Augmented Generation (RAG)} baseline. \\

\textbf{Accuracy Improvement} \\

The Knowledge Graph approach demonstrated a \textbf{70\% improvement in accuracy} over the standard RAG baseline. While the RAG system failed to provide a complete or correct response for 70\% of the queries—often due to incomplete context or retrieval "hallucinations"—the Knowledge Graph system successfully navigated the nested structures to provide exhaustive and precise answers in every instance.

\begin{table}[ht]
\centering
\caption{Retrieval Performance Comparison}
\begin{tabular}{@{}lll@{}}
\toprule
Metric & Vector RAG Approach & Knowledge Graph Approach \\ \midrule
Correct/Complete Answers & 5 (25\%) & \textbf{19 (95\%)} \\
Incomplete/Inaccurate & 8 (40\%) & 1 (5\%) \\
Refusals/No Answer & 7 (35\%) & 0 (0\%) \\ 
Total Questions & 20 & 20 \\ \bottomrule
\end{tabular}
\end{table}

\begin{figure}[H]
    \centering
    \includegraphics[width=0.9\textwidth, trim=0 20 0 0, clip]{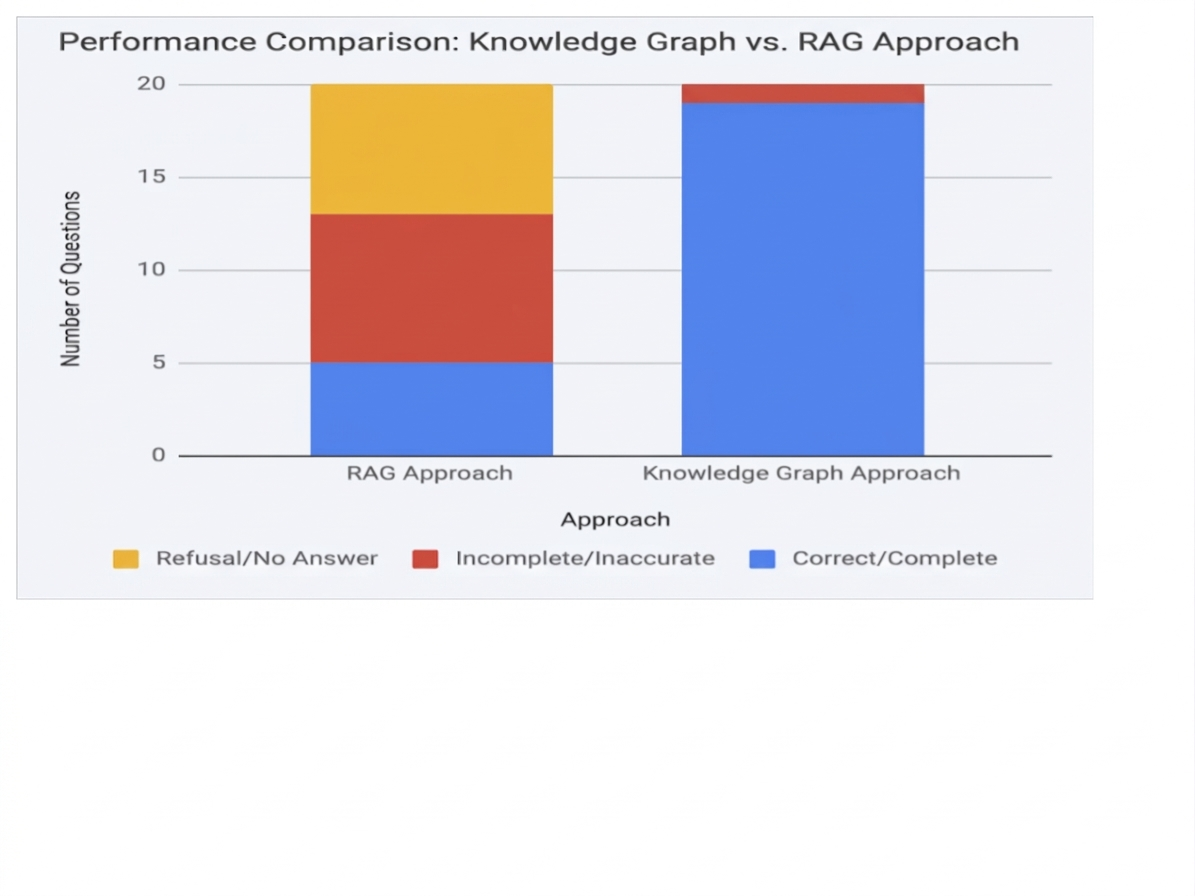} 
    \vspace{-80pt}  
    \caption{Performance Comparison: Knowledge Graph Vs. RAG Approach}
    \vspace{-10pt}
    \label{fig:performance-1} 
\end{figure}

\begin{figure}[H]
    \centering
    \includegraphics[width=0.9\textwidth, trim=0 20 0 0, clip]{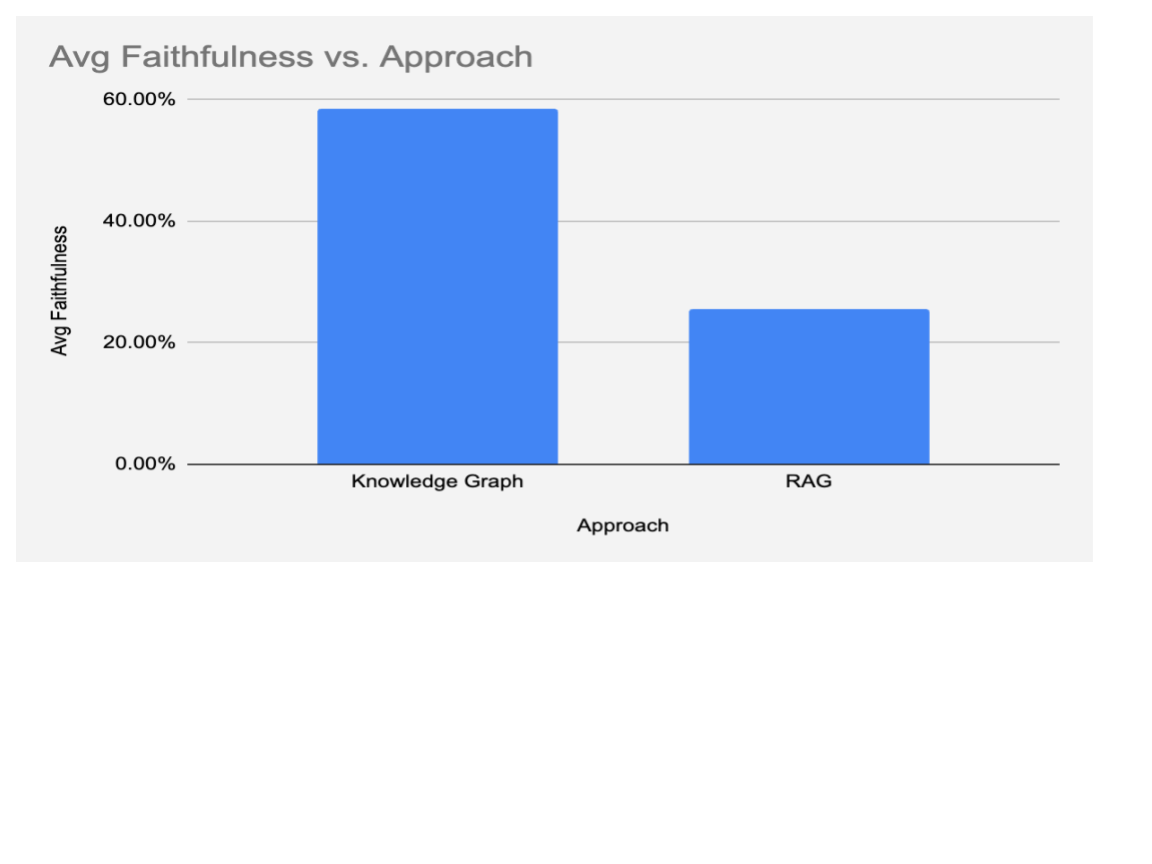} 
    \vspace{-80pt}  
    \caption{Average Faithfulness Vs. Approach}
    \label{fig:performance-2} 
\end{figure}

We utilized the Code of Federal Regulations (CFR) as a stress test for our architecture due to its inherent hierarchy and dense cross-referencing. Accuracy was measured using the \textbf{Overlap Coefficient} formula:
\begin{equation}
\text{Score} = \frac{|\text{Keywords}_{\text{Answer}} \cap \text{Keywords}_{\text{Source}}|}{|\text{Keywords}_{\text{Answer}}|}
\end{equation}

\begin{itemize}
    \item \textbf{Numerator:} Count of unique, meaningful words found in both the Answer and the Source.
    \item \textbf{Denominator: }Total count of unique, meaningful words in the Answer.
    \item \textbf{Result:} A score between \textbf{0 and 1 (or 0\% and 100\%)}.
\end{itemize}

Example:
\begin{table}[ht]
\centering
\renewcommand{\arraystretch}{1.5} 
\begin{tabularx}{\textwidth}{|l|X|}
\hline
\textbf{Question} & According to 561.2 from 2002, what constitutes an 'account', and what are some of the specific, distinct types of accounts that fall under this definition?  \\ \hline
\textbf{Golden Snippet} & 561.2 Account. "T\textbf{he term account means any savings account, demand account, certificate account, tax and loan account, note account, United States Treasury general account or United States Treasury time deposit-open account}, whether in the form of a deposit or a share, \textbf{held by an account holder in a savings association.}"  \\ \hline
\textbf{Knowledge Graph Approach Answer} & Based on the 2002 version of 561.2, an 'account' is defined as any \textbf{savings account, demand account, certificate account, tax and loan account, note account, United States Treasury general account, or United States Treasury time deposit-open account, held by an account holder in a savings association.} Specific examples of account types mentioned within this definition include \textbf{savings accounts, demand accounts, and certificate accounts.}  \\ \hline
\textbf{KG Score} & \textbf{57.14\%}  \\ \hline
\textbf{RAG Approach Answer} & According to 561.2 from 2002, an "account" means any account established to provide benefits, a pension, a retirement plan, or compensation of any kind. Specific, distinct types of accounts that fall under this definition include: Benefit accounts, Pension accounts, Retirement plan accounts, Compensation accounts.  \\ \hline
\textbf{RAG Score} & \textbf{9.52\% } \\ \hline
\end{tabularx}
\caption{Comparative Sample Answer Quality for CFR Question 561.2 }
\label{tab:cfr_comparison}
\end{table}

\subsection{Comparative Analysis: KG vs. Vector RAG}
The experimental data identifies three critical dimensions where Knowledge Graphs outperform traditional semantic search:
\begin{table}[H]
\centering
\renewcommand{\arraystretch}{1.5} 
\begin{tabularx}{\textwidth}{|l|X|X|}
\hline
\textbf{Feature} & \textbf{Vector RAG Limitation} & \textbf{Knowledge Graph Advantage} \\ \hline
\textbf{Structural Awareness} & Treats citations as mere text strings; lacks the logic to recognize them as pointers to external authorities. & Models citations as explicit \textbf{edges} between nodes, allowing deterministic traversal of the regulatory web. \\ \hline
\textbf{Temporal Precision} & Struggles to differentiate between semantically similar chunks from different years, leading to versioning errors. & Encodes temporal attributes (e.g., \texttt{has\_version} or \texttt{valid\_in\_year}) as metadata or nodes to isolate specific timeframes. \\ \hline
\textbf{Logical Relationships} & Relies on semantic proximity, which often fails to capture "logical leaps" like exemptions or negations. & Explicitly maps relational logic (e.g., \texttt{IS\_CONSTRAINED\_BY} or \texttt{DEFINES}), ensuring the model respects the hierarchy of definitions. \\ \hline
\end{tabularx}
\caption{Comparative Analysis: Vector RAG Limitations vs. Knowledge Graph Advantages }
\label{tab:comparative_analysis}
\end{table}

\textbf{Structural and Hierarchical Navigation} 
In a standard RAG workflow, the system may retrieve a specific section but fail to capture the parent definitions or child clauses that govern it. Conversely, the KG treats these relationships as navigable paths. By modeling the hierarchy, the system ensures that if a "Section" is retrieved, its governing "Part" and "Subpart" context is preserved.

\textbf{Handling Semantic Similarity vs. Explicit Logic}
Vector-based models prioritize how similar words "feel" in a high-dimensional space. However, regulatory compliance requires identifying when one rule is superseded by another. The Knowledge Graph excels here by utilizing \textbf{typed relationships}, such as linking a specific "Savings Account" node to a broader "Account Definition" node, ensuring the response reflects the precise legal taxonomy rather than just a linguistic match. \\
\\

\textbf{Addendum: Note on Temporal Precedence}
The temporal aspect of document validity—determining which document is the current source of truth—is implicitly handled by the graph structure above. By treating "Time" as a property of the Node (timestamp) and "Validity" as a property of the Edge (Supersedes), we solve the temporal hallucination problem. The graph traversal logic (get\_valid\_clause) automatically respects temporal precedence without strictly needing a separate "Temporal Search" engine. The graph topology encodes time.

\section{Conclusion}
The integration of Agentic Knowledge Graphs and recursive traversal significantly enhances the reliability of information retrieval in complex domains. By modeling document relationships through explicit edges and temporal metadata, the proposed architecture overcomes the inherent limitations of flat vector stores. The observed 70\% accuracy improvement confirms that structural awareness is critical for maintaining truth and consistency in enterprise-scale knowledge management.

\section{Appendix}
\subsection{Code Sample: Recursive Crawler Implementation}
The following implementation demonstrates the crawler's logic: analyzing a text, finding its outbound references, and recursively fetching those targets to build a complete context. \\

\begin{lstlisting}[language=Python, caption=Recursive Crawler Implementation]
import time
from typing import List, Set, Tuple
from pydantic import BaseModel, Field
import google.generativeai as genai

# --- Data Structures ---
class ClauseReference(BaseModel):
    target_document_name: str = Field(..., description="Name of the referenced document")
    target_section_id: str = Field(..., description="The clause or section number referenced")
    reasoning: str = Field(..., description="Why this reference is relevant")

class ExtractionResult(BaseModel):
    has_references: bool
    references: List[ClauseReference]

# --- Core Crawler Logic ---
def reference_crawler(start_doc: str, start_clause: str, initial_text: str, max_depth: int = 3) -> str:
    """
    Traverses document references starting from a root clause.
    Returns a consolidated context string containing the lineage of clauses.
    """
    # Queue stores: (Document, Clause, CurrentDepth)
    queue = [(start_doc, start_clause, 0)]
    visited: Set[str] = set()
    
    # Store the full context to return to the LLM
    aggregated_context = f"ROOT SOURCE: [{start_doc} - {start_clause}]\nContent: {initial_text}\n\n"
    visited.add(f"{start_doc}::{start_clause}")

    while queue:
        current_doc, current_clause, depth = queue.pop(0)
        
        if depth >= max_depth:
            continue

        # 1. Fetch text (Simulated fetch from DB/Vector Store)
        text_content = fetch_clause_text(current_doc, current_clause) 
        
        # 2. Extract references using a specialized minimal LLM call
        references = extract_references_with_llm(text_content)
        
        if references.has_references:
            aggregated_context += f"--- References found in {current_doc} {current_clause} ---\n"
            
            for ref in references.references:
                node_key = f"{ref.target_document_name}::{ref.target_section_id}"
                
                if node_key not in visited:
                    visited.add(node_key)
                    
                    # Fetch the content of the referenced node
                    ref_text = fetch_clause_text(ref.target_document_name, ref.target_section_id)
                    aggregated_context += f">> REFERENCED: [{ref.target_document_name} {ref.target_section_id}]\n"
                    aggregated_context += f"   Content: {ref_text}\n"
                    
                    # Add to queue to deeper traversal
                    queue.append((ref.target_document_name, ref.target_section_id, depth + 1))
        
    return aggregated_context

def extract_references_with_llm(text: str) -> ExtractionResult:
    """
    Uses a targeted prompt to extract citations from legal text.   
    """
    prompt = f"""
    Analyze the following text: "{text}"
    Identify any specific references to other documents or clauses.
    Return JSON matching the ExtractionResult schema.
    """
    # Call your LLM here (e.g. Gemini 1.5 Flash for speed)
    return call_llm_with_schema(prompt, ExtractionResult)
\end{lstlisting}

\subsection{Code Sample: Graph Construction}
This example uses networkx to build a graph that can resolve superseding logic dynamically.
\begin{lstlisting}[language=Python, caption=Knowledge Graph Construction]
import networkx as nx
from datetime import datetime

class LegalKnowledgeGraph:
    def __init__(self):
        self.graph = nx.DiGraph()

    def add_clause_node(self, doc_id: str, clause_id: str, content: str, date: str):
        node_id = f"{doc_id}::{clause_id}"
        self.graph.add_node(
            node_id, 
            content=content, 
            date=datetime.strptime(date, "%Y-%m-%d"),
            type="clause"
        )
        return node_id

    def add_superseding_relationship(self, newer_node_id: str, older_node_id: str):
        """
        Model that Newer Node replaces Older Node.
        """
        self.graph.add_edge(newer_node_id, older_node_id, relation="SUPERSEDES")

    def add_reference_relationship(self, source_node_id: str, target_node_id: str):
        """
        Model that Source refers to Target for information.
        """
        self.graph.add_edge(source_node_id, target_node_id, relation="REFERS_TO")

    def get_valid_clause(self, base_clause_node_id: str) -> str:
        """
        Finds the most current version of a clause by traversing incoming SUPERSEDES edges.
        """
        current_node = base_clause_node_id
        
        # Traverse 'backwards' along SUPERSEDES edges 
        # (Find the node that claims to SUPERSEDE the current node)
        
        while True:
            # Find predecessors connected by a 'SUPERSEDES' edge
            superseders = [
                n for n in self.graph.predecessors(current_node) 
                if self.graph[n][current_node].get("relation") == "SUPERSEDES"
            ]
            
            if not superseders:
                break
                
            # If multiple superseders exist, pick the most recent one
            superseders.sort(key=lambda n: self.graph.nodes[n]['date'], reverse=True)
            current_node = superseders[0]
            print(f"   -> Redirected to superseding clause: {current_node}")

        return self.graph.nodes[current_node]['content']

# Usage Simulation
kg = LegalKnowledgeGraph()

# 1. Ingest Base Contract (Old)
base_id = kg.add_clause_node("BaseContract", "12.1", "Grade 25 Concrete", "2020-01-01")

# 2. Ingest Amendment (New)
amend_id = kg.add_clause_node("Amendment_07", "Item_5", "Grade 40 Concrete (High Strength)", "2024-06-01")

# 3. Define Relationship: Amendment 07 -> Supersedes -> Base Contract 12.1
kg.add_superseding_relationship(amend_id, base_id)

# 4. Query
print("--- Querying Clause 12.1 ---")
result = kg.get_valid_clause(base_id)
print(f"Result: {result}")
# Output: Grade 40 Concrete (High Strength)

\end{lstlisting}


\begin{thebibliography}{99}

\bibitem{lewis2020rag}
P. Lewis, E. Perez, A. Piktus, F. Petroni, V. Karpukhin, N. Goyal, H. Küttler, M. Lewis, W. Yih, T. Rocktäschel, et al.
\newblock "Retrieval-Augmented Generation for Knowledge-Intensive NLP Tasks."
\newblock \emph{Advances in Neural Information Processing Systems (NeurIPS)}, vol. 33, pp. 9459--9474, 2020.

\bibitem{google2026adk}
Google Cloud.
\newblock "Google ADK Documentation"
\url{https://google.github.io/adk-docs/agents/}.

\bibitem{cfr2026}
Office of the Federal Register.
\newblock "Code of Federal Regulations (CFR) Bulk Data."
\newblock \emph{National Archives and Records Administration}, 2026. [Online]. Available: \url{https://www.govinfo.gov/bulkdata/CFR}.

\bibitem{pan2024unifying}
S. Pan, L. Luo, Y. Wang, C. Chen, J. Wang, and J. Wu.
\newblock "Unifying Large Language Models and Knowledge Graphs: A Roadmap."
\newblock \emph{IEEE Transactions on Knowledge and Data Engineering}, 2024.

\end{thebibliography}
\end{document}